\documentclass[article,9pt,letterpaper]{revtex4}
\usepackage[dvips]{graphicx,color}
\usepackage{epsfig,epstopdf,ifpdf}
\usepackage{latexsym,slashed}
\usepackage{subfigure,amsmath,amssymb,amsfonts}
\begin{document}
\title {Asymptotic Methods in Non linear dynamics} 

\author{R Dutta  \footnote{\it rdutt$@$orca.st.usm.edu} }
\affiliation { Department of Mathematics, The Ohio State University  }
\maketitle 

\section{Abstract} 
This paper features and elaborates  recent developments and modifications  in asymptotic techniques 
in solving differential equation in non linear dynamics. 
These methods are proved to be powerful  to solve  weakly as well as  strongly non linear cases. 
Obtained   approximate analytical solutions   are valid for the whole solution domain.
In this paper,  limitations of traditional perturbation methods are illustrated with various modified techniques. 
Mathematical tools such as  variational approach, homotopy and iteration technique are discussed  to 
solve various problems efficiently. 
Asymptotic methods such as Variational Method, modified Lindstedt-Poincare method, Linearized perturbation method, 
Parameter Expansion method,  Homotopy Perturbation method 
 and  Perturbation-Iteration  methods(singular and non singular cases)  have been discussed in various situations. 
 Main emphasis is given  on Singular perturbation method and WKB method  in various numerical problems.  \\
\vskip 1pt 
keywords: analytical solution, nonlinear equation, perturbation method, variational theory, homotopy perturbation, 
variational iteration method, correctional functional.  \\

\section{Introduction} 
 With rapid development of research in  non linear science, there appears ever increasing interest of scientists and 
 engineers in 
 asymptotic techniques to solve  non linear problems, specially non linear differential equations. 
  First order non linear partial equations model non linear waves, gas dynamics,
  water wave, elastodynamics, chemical reaction, transport of pollutants, flood wave in river, chromatography, 
 traffic flow and wide range of biological and ecological systems.  One of the most important non linear phenomena
  without non linear counterpart is the breakdown of solution in finite time resulting in  formation of 
  discontinuous shock wave. A striking example is the supersonic boom produced by an airplane. 
 In linear wave equation, the signal propagates along with the characteristics.   
 But in non linear case, the characteristics can cross each other precipitating an  onset of shock. 
 The characterization of shock dynamics requires additional physical information.  
 Parabolic second order differential equations govern non linear diffusion processes including thermodynamics, 
 chemical reaction, dispersion of pollutants and population dynamics. The simplest and most well understood 
 is Burger equation which can surprisingly be linearized by transforming it to heat equation. 
 In the limit of diffusion or viscosity tending to zero, solution of Burger equation tends to shock wave solution 
 to the limiting first order dispersion less equation and thus provides alternative
  mechanism for unravelling shock dynamics.   Second order non linear equations arise in fluid dynamics,  water wave, 
   Blasius layer theory  i.e  Falkner Skan equation,
  heat wave and heat equation that include 
  wide variety of equations such as 
  autonomous equation, Emden-Fowler equation , Navier Stokes equation. 
  Third order partial differential equations arise in  study of dispersive wave motion including water wave, 
  plasma wave, wave in elastic medium( very suitable in arterial blood flow studies).  
   It is very easy to find solution in linear system numerically but becomes
  very difficult  to find solution for non linear equation analytically or numerically. Because  almost all discretized 
   methods or numerical simulation use
  iteration method technique in solving equation which is very sensitive to initial solutions \cite{gu, bao, chen, liu2}.  
  It implies that  it is very difficult to find solution in convergence in presence of non linearity, often impossible in 
  case of strong one.
  The  problem that we encounter 
  quite often is that amplitude of  non linear oscillation which depends on initial parameter becomes lost during 
  numerical simulation. 
   Analytical asymptotic approaches to solve non linear equations include 
  Non Perturbative method \cite{dela}, Adomian Decomposition method \cite{abbas, ado, danaf, waz} , 
  Modified Lindstedt-Poincare method  \cite{cheung, he9}, Variational Iteration method \cite{he1, he7, odibat} , 
  Energy Balance method \cite{he8}, 
  F-expansion method \cite{ren, wang1} and 
   Perturbative  methods \cite{awr, bellman, dutta1, micken, nayfeh1}  are most common approximate
  methods in studying non linear mathematical models arising in physics and engineering that are
   constantly being developed 
  or applied to more complex non linear systems. 
  A major limitation of these methods is the restriction of small parameter which
  makes the solution valid for weakly non linear systems mostly,  very difficult to 
  find solution in strongly non linear case. 
   Alternative perturbative techniques such as Linearized Perturbation method \cite{he1},  
  Modified Linstedt-Poincare method \cite{cheung},  Homotopy Perturbation method \cite{he2} and
   Multiple Scale Lindstedt-Poincare method \cite{pakdemirli} 
  are being developed and analyzed in various systems. 
   Alternative attempt to validate solutions for strongly non linear systems is  the use of Iteration - Perturbation technique 
  \cite{marinca, yu} . 
  On the other hand, Stochastic methods[non asymptotic method]  \cite{dutta2}are being applied numerically to solve qualitative 
  and quantitative 
  properties of the solution for various design of engineering and Industrial processes. 
  Non linear equations \cite{high} assumes  the system to be iterative and  these methods prove
   to be 
  powerful tool to solve non linear dynamics undergoing random fluctuation in  physical
   parameters through perturbative technique. 
 Many of these methods 
 or numerical simulations apply iteration technique  to find their numerical solutions of non linear problems 
and nearly all iterative solutions are sensitive to initial solutions. 
Convergence of these solutions is the main difficulty.
In addition, natural frequency of non linear oscillation  depends on initial condition that means  amplitude of the oscillation 
 will be lost during the procedure of numerical simulation.  
 Variation Iteration method (VIM) based on the use of restricted variations and correction functionals has found 
  wide applications for  solution of non linear ordinary and partial differential equations. 
  Almost all perturbative  methods are based on assumption with the existence of a small parameter in  equation. 
 This parameter assumption greatly restricts applications of perturbation techniques, a well known problem in solving 
 an overwhelming majority of non linear problems. 
 It is even more difficult to determine this small parameter which requires special technique and often
  arises from  understanding of  the physical system.  An inappropriate choice of small parameter 
  results in bad effect (sometimes serious effect) on the solution. 
   In presence of suitable parameter value,  approximate solution obtained using  perturbative methods are 
  valid only for small values of the parameter in most cases. For example, approximations solved by multiple scaling 
  are uniformly valid as long as specific system parameter is small. However, we can not rely on 
  approximation fully because there is no criterion on how small the parameter should be. 
  Thus numerical result or experimental value of the approximations is necessary.  
  We discuss various analytical and numerical methods to solve various non linear situations . First , we will discuss 
  variational approaches such as Ritz method, Soliton solution, Bifurcation , Variational Iteration method. Under modified 
  expanding parameter method, Lindstedt - Poincare method(modified form) , 
  Parametrized Perturbation method and Homotopy perturbation method with various 
  equations under non linear dynamics.  Finally, we will discuss about Singular perturbation method  and 
  WKB methods implemented  
  in various non linear situations.
  If we consider  following  equations \\
  \begin{equation}
  \begin{split} 
  u^{\prime} + u - {\epsilon} = 0  \cdots \text{with } u(0)=1 \\
  u^{\prime} -u - {\epsilon} = 0 \cdots \text{with} u(0)=1 
  \end{split} 
  \end{equation}
  The solution of the first of equation (1) is 
 \[   u(t) = {\epsilon} + {(1-\epsilon)}e^{-t}   \]  
 while the unperturbed solution is \[ u_o(t) = e^{-t} \] . In case $ \epsilon \ll   1 $, 
 the perturbation solution leads to high accuracy for 
\[    \mid u(t) - u_0(t) \mid \le \epsilon   \]
 By similar analysis of the second   of  equation (1) we have , 

\[  \mid   u(t) - u_0(t) \mid = {\epsilon} \mid (1-e^{-t} ) \mid   \]

 It is obvious that $u_0$ can never be considered as approximate solution of equation (1) for $ t \gg  1$ . 
 So, normal perturbative method is not valid even $ \epsilon \ll  1 $. \\
 This indicates that we should introduce new developed method when normal 
 perturbative  method fails.    There exist analytical asymptotic approaches such as non 
 perturbative method \cite{dela} , Weighted linearized method \cite{agrawal, ramos} , 
  Adomian decomposition method \cite{abbas, danaf, waz} , 
 Modified Lindstedt-Poincare method \cite{cheung, he9} , Variational Iteration method  \cite{he7,he1} , 
 Energy Balance Method \cite{ he8} , 
 tanh method \cite{ba} ,
 F-expansion method \cite{ren} and so on. Some new perturbation methods  such as Artificial Parameter Method \cite{he9} 
 and $\delta$ method \cite{awr,liu}, Perturbation Incremental method \cite{cheung}, 
 Homotopy perturbation method \cite{he5} are
  being proposed to solve these 
 systems which  does not depend on small parameter. \\
 Recent study reveals that numerical technique can also be powerfully 
 applied to  perturbative  methods. 
 A wide literature dealing with problem of approximate solutions in non linear equations with various methods have been 
 reviewed. But comprehensive review of asymptotic methods for non linear equations still requires more research.   There exist 
 some alternative analytical asymptotic approaches such as non perturbative method to probe non linear equations such as 
 Modified linearization method, Adomian decomposition method \cite{abas,ado}, Lindstedt-Poincare method \cite{cheung}, 
 Variational Iteration method \cite{bildik}, 
  Parameter Expanding method \cite{he12} and so on. In parameter expanding perturbative method,   parameter can be expanded as
   power series of $\epsilon$ which can be a small parameter,  an artificial parameter or book keeping parameter. 
   Perturbative method proves to be useful tool 
 to analyze stochastic structure  of an chaotic system very well. 
 In this work, we limit to non linear partial differential equation (NPDE) equations  because  most PDE can be converted 
 to ODE by linear 
 transformation . We start with  KdV equation, 
 \begin{equation} 
{  \partial_t } {\phi} + b {\phi} {\partial_x }{\phi}  + {\epsilon} {\partial_{xxx}}{\phi}  = 0 
 \end{equation} 
 KdV is a non linear, dispersive PDE for a function $\phi$ of two variable , space and time. 
 If  we  assume that $ {\phi} (x,t) = {\phi} (\xi) = {\phi} (x-ct) $. where c is  wave speed. 
 Substituting above into wave equation, we get 
 \begin{equation}
 - cu_{\xi} = + u u_{\xi} + {\epsilon} u_{\xi \xi \xi } 
 \end{equation} 
 Integrating Eq (), we have an Ordinary differential equation 
 \begin{equation} 
 -cu  + \cfrac{1}{2} u^2 + {\epsilon}u_{\xi \xi } = A
\end{equation} 
with A as constant of integration. Interpreting $\xi$ as virtual time variable, this signifies that $\phi$  satisfies Newton
 motion in cubic potential.  
If $\epsilon$ is  adjusted so that potential $V(\phi)$ has local minimum for $\phi=0$ 
function , there exists solution in which  $\phi(\xi)$ starts at virtual time
and eventually reaches to local minima.  The characteristic shape of the function is called Solitary wave used in solution 
of soliton equation.  More precisely, the solution is 
\begin{equation} 
{\phi}(x,t) = \cfrac{1}{2} c sech^2 \left[ \cfrac{\sqrt{c}}{2} (x-ct - a) \right] 
\end{equation}
This describes right moving soliton.  KdV equation has infinitely integrals of motion. 
In next section, variational approach to soliton equation , bifurcation, limit cycle and periodic solution 
of non linear equations are illustrated including Ritz method, energy method , Iterative variational method. 
A recent study \cite{he6} reveals that the numerical technique can also be applied powerfully to perturbation method. 

\section{Variational Approach}
\subsection{Ritz method} 
Variational methods \cite{fin,liu3, zhang1} such as Raleigh-Ritz and Galerkin methods have been popular tools for 
non linear analysis. 
Compared to other methods, variational method provides physical insight into the nature of the solution. 
Ritz method is a direct method to find an approximate solution for boundary value problem.  It is exactly the finite 
element method 
used to compute Eigen vector and Eigen values of a hamiltonian system. Rayley - Ritz method allows  computation
of ritz pairs  ( $ \tilde{\lambda_i}, \tilde{x_i} $) which approximates solution to Eigen value problem. 
In this method, the problem must be stated in a variational form as a minimization problem and solution is approximated 
by a finite linear combination of all possible trial functions.  
As an illustration,
we  apply Ritz method to Lambert equation 
 \begin{equation} 
 \ddot{y} (x) + \cfrac{k^2}{n} y(x) - (1-n) \cfrac{ {\dot{y} (x) }^2} {y(x)}  =0 
 \end{equation}
and arrived at variational formulation with energy integral in Hamiltonian system  as
\begin{equation}
J(y) = \cfrac{1}{2} \int ( - n^2 y^{2n-2}  {\dot{y}}^2 + k^2 y^{2n} ) dt 
\end{equation} 
By transformation  $ z= y^n$ , we obtain 
\begin{equation}
J(z) = \cfrac{1}{2} \int (- {\dot{z}}^2 + k^2 z^2 ) dt 
\end{equation} 
which leads to  linear Euler Lagrange equation
\begin{equation}
\ddot{z}  + k^2 z = 0 
\end{equation} whose solution is of the form 
\begin{equation}
y = {( C \ cos kx + D \ sin kx) }^{1/n}
\end{equation} 
Considering non linear wave equation i.e kdV equation 
\begin{equation}
\cfrac{\partial u}{\partial t} - 6 u \cfrac{\partial u}{\partial x} + \cfrac{\partial^3 u}{\partial x^3} = 0 
\end{equation}
We seek traveling wave solution of following form 
\begin{equation}
u(x,t) = u(\xi) 
\end{equation} with $ \xi = x-ct $. Substituting the solution into differential equation we obtain 
\begin{equation}
-c \dot{u} - 6 u \dot{u} + \ddot{u} = 0 
\end{equation} 
Here prime denotes differential with respect to $\xi$.  This is soliton equation.  And after integrating the above equation,
 one obtains 
\begin{equation}
-cu -3u^2 + \ddot{u}  = 0
\end{equation}
By semi inverse method \cite{taniuti} , following variational formulations is obtained 
\begin{equation}
J = {\int_0}^{\infty} \left( \cfrac{1}{2} c u^2 + u^3 + \cfrac{1}{2} {( \cfrac {du}{d \xi} )}^2  \right) d {\xi} 
\end{equation}
Implementing  Ritz method, we try to find solution of the form
\begin{equation}
u = p sech^2 ( q \xi) 
\end{equation}
Upon substituting in the differential equation, we finally obtain solution as 
\begin{equation}
u = - \cfrac{c}{2} \ sech^2 \sqrt{\cfrac{c}{4}}(x-ct- \xi_0) 
\end{equation}
which is exact solitary wave solution of KdV equation. 
We now  focus on Ritz method in bifurcation problem. 
Bifurcation arises in various non linear problems \cite{zhang2}.   One can apply Variational method  to search for 
 dual solution of a nonlinear equation if it exists. Ritz method  method helps to identify bifurcation point. 
Considering Bratu equation, we apply Ritz method to nonlinear Bratu equation of the form
\begin{equation}
\ddot{u} + {\lambda} e^u = 0 
\end{equation}
with boundary condition $ u(0)=u(1)=0$ 
Bratu type equation arises in many physical \& Engineering problems  such as chemical reaction, radiative heat transfer
,  nanotechnology and universe expansion model.  As an example, the above equation is typically obtained  from solid 
fuel ignition model in thermal combustion which has exact solution. 
The above problem can be solved with high accuracy using variational method.  From above equation, one can
 obtain variational formulation as 
\begin{equation}
J(u) = {\int_0}^1 { \cfrac{1}{2} {\dot{u}}^2 - {\lambda} e^u } dx
\end{equation}
We can choose trial function as 
\begin{equation}
u = A x (1-x) 
\end{equation}
Here A is unknown constant to be determined. Substituting u into variational formulation(integral form), one can obtain
\begin{equation}
J(A) = {\int_0}^1 { \cfrac{1}{2} A^2 {(1-2x)}^2 - {\lambda} e^{Ax(1-x)} } dx 
\end{equation}
Using minimization principle with respect to A , this yields 
\begin{equation} 
\cfrac{\partial J}{\partial A} = {\int_0}^1  {A{(1-2x)}^2 - {\lambda} x (1-x) e^{Ax(1-x)} } dx = 0
\end{equation}  
One can find out value of critical value for unknown constant A numerically showing that 
Bratu equation has two solutions which is the indication bifurcation phenomena for certain A. 

\subsection{Limit Cycle Method}
In mathematics, in the study of dynamical system with two dimensional phase space, a limit cycle is a
 closed trajectory that spirals into it either as time approaches negative infinity or positive infinity.  Such behavior
 is exhibited in some non linear systems.  Limit cycle can be used to model the behavior of many real world 
 oscillatory systems.  To illustrate the basic idea of the method, one can consider non linear oscillation as
 \begin{equation}
 \ddot{x} + x + {\epsilon} f(x, \dot{x}) = 0 
 \end{equation}of this equation 
 Limit cycle  can be determined in general form as 
 \begin{equation}
 x = a + b(t) \ cos \omega t + \sum c_n cos (n \omega t ) + d_n \ sin (n \omega t )
 \end{equation}
 in terms of Fourier Series \cite{he11}.   Assuming amplitude as varying function of time 
 that is \[   a(t) \approx A e^{\alpha t }  \] , one can calculate ${ \alpha} $ for Van der Pols equation
based on limit cycle method as
 \begin{equation}
 \alpha = \cfrac{ {\epsilon}^2 (A^2 -4) {\omega} {\pi}} {4( 2 {\pi} {\omega} - {\epsilon} A^2} 
 \end{equation}
 In case of electro spin problem,  any nth order solution can be written as
 \begin{equation} 
 u_{n+1} (z) = u_n(z) + {\int_0}^z  \left[   (s-z) \left(   \cfrac{\partial^2 u_n(s) }{\partial s^2} + f( u_n(s) \right)   \right]  ds 
 \end{equation} 
 Limit cycle method helps to study  if the system exhibits any oscillation or not. 
He et al \cite{he11} implemented limit cycle  method for non linear oscillator system which represents 
 various engineering nano mechanical problem.  
 
 \section{Variational Iteration Method} 
The variational iteration method \cite{he1,he11,odibat} has been proved  to solve 
 a large class of non linear problems  very effectively and accurately with approximations 
converging rapidly to accurate solutions.  Many authors found that shortcoming of Adomian method can be improved using 
Variational Iteration method \cite{sweilam, odibat, momani}.  Illustrating basic idea of variational Iteration method, 
we consider  equation of  general non linear system as 
\begin{equation} 
 Lu + Nu = g(x)
\end{equation}
where L is the linear operator and N is the non linear one. According to variational iteration method, we can construct following iteration 
formulation 
\begin{equation}
u_{n+1}(x) = u_n(x) + {\int_0}^x {\lambda}( Lu_n(s) + N \tilde{u_n(s)}  - g(s) ) ds 
\end{equation}
 where $\lambda$ is the Lagrangian multiplier which can be found implementing using variational theory i,e $ \delta u_n =0 $ .    
 
\subsection{Variational method to Duffing Equation}
Duffing equation is a classic  example in non linear dynamics to illustrate numerical procedure. Duffing equation is an example of a 
dynamical system that exhibits chaotic behavior. This 
equation is a non linear second order differential equation used to model certain 
damped and force  driven oscillator given by 
\begin{equation} 
\ddot{u}  + {\delta} \dot{u}  + {\alpha} u + {\beta} u^3  = {\gamma} cos ( \omega t) 
\end{equation}
with initial condition  $u_(t=0) =0 $  , $ \dot{u(t=0)} =0 $ 
The equation describes the motion of a damped oscillator with more complicated potential than in 
simple harmonic motion.  Suppose that the angular frequency of the system  is $\omega$, 
then we have the following homogeneous linearized equation
\begin{equation} 
\ddot{u}  + {\omega}^2 u = 0 
\end{equation} 
so that we can write above equation () in following form 
\begin{equation}
\ddot{u}  + {\omega}^2 u + g(u) = 0 
\end{equation}
where $ g(u) = {\delta}\dot{u} + {\beta} u^3 - {\gamma} \ cos ( \omega t) $ 
Implementing variational iteration method, we can construct functional as 
\begin{equation} 
u_{n+1} (t) = u_n (t) + {\int_0}^t {\lambda} { \ddot{u_n}(\tau)  + {\omega}^2 u_n(\tau) + \tilde{g(u_n)} } d \tau
\end{equation} 
where $ \tilde{g(u_n)} $ is considered as restricted variation in variational theory i.e $ {\delta} \tilde{g} =0 $. 
Calculating variation with respect to $u_n$, and using restricted variation theory , we have following stationary conditions
\begin{equation} 
\begin{split}
\ddot{\tau} + {\omega}^2 {\lambda}(\tau) =0 \\
\lambda (\tau)_{\tau=t} = 0 \\
\end{split}
\end{equation}
The multiplier can then be identified as 
\begin{equation}
\lambda = \cfrac{1}{\omega} \ sin {\omega} 
\end{equation}
Substituting $lambda$ value into iteration equation , we obtain
\begin{equation}
u_{n+1} (t) = u_n(t) + \cfrac{1}{\omega} \ sin( \tau-t) 
\end{equation}
Substituting this into iteration formula, we get
\begin{equation} 
u_{n+1} (t) = u_n(t) + \cfrac{1}{\omega}  {\int_0}^t \ sin( \omega(\tau-t) { \ddot{u_n}(\tau) + u_n(\tau) + {\beta}{u_n}^3 (\tau) + f(\tau) } d {\tau} 
\end{equation}
Assuming initial approximate solution of the form 
\begin{equation}
u_0(t) = A \ cos ({\omega} t )
\end{equation} 
leads to calculation of residual function as 
\begin{equation}
R_0(t) = \left( 1 -{\omega}^2 + \cfrac{3}{4} A^2 \right) A \ cos {\omega} t  + \cfrac{1}{4} \ cos {\omega} t + \cfrac{1}{4} \ cos 3 {\omega} t 
\end{equation}
And using iteration formulation, one can calculate higher order also
\begin{equation}
u_1(t) = A \cos {\omega} t + \cfrac{\epsilon A^3 } {32  \omega^2 } ( \ cos 3 {\omega}t - \cos {\omega}  t ) 
\end{equation}
\subsection{Variational Iteration approach to Bratu equation} 
Due to various engineering and physical  importance, Bratu problem has been studied extensively.  
It arises in a wide variety of physical applications ranging 
from chemical reaction theory to radiative heat transfer and nanotechnology. 
Because of  simplicity , the equation is widely used as
benchmark tool for 
numerical method.
The  problem can be easily solved with high accuracy by the variational iteration method.
 Furthermore, two branches of the solutions can be simultaneously 
determined and bifurcation point can be identified with ease, if there is any. 
The classical Bratu problem is 
\begin{equation}
\Delta u + {\lambda} e^u = 0 
\end{equation}
with planer version as
\begin{equation}
\ddot{u(x)} + {\lambda} e^{u(x) } = 0
\end{equation}
with $ 0 \le x \le 1 $ and with Dirichlet boundary condition  $ u(0) =0 $ and $ u(1) =0 $ 
This equation is used in combustion modeling. 
It is easy to establish a variational formulation for Eq () which reads 
\begin{equation}
 J(u) = {\int_0}^1 { \cfrac{1}{2} {u^{\prime}}^2 - {\lambda} e^{u} } dx 
\end{equation}
  Variational iteration method  can be successfully applied to bratu type equation to construct 
 correction functional 
\begin{equation} 
u_{n+1}(x) = {u_n}(x) +  {\int_0}^x [ \cfrac{d {u_n}^2(s)} { d s^2} + {\alpha} e^{\tilde{u_n}(s) } ] ds 
\end{equation} 
$ n \ge 0$ with restricted variation  $ \delta \tilde{u_n} = 0 $ to find multiplier value. Using  Iteration Variation method gives,
\begin{equation}
u_{n+1}(x) = u_n(x) + {\int_0}^x (s-x) [ \cfrac{d {u_n}^2}{d s^2}  + {\alpha} e^{u_n(s)} ] ds 
\end{equation}
Batiha et al \cite{batiha} showed using numerical results that variation iteration method is
 more effective than Adomian decomposition method.

\section{Modified Lindstedt Poincare Method} 
In perturbation theory, Lindstedt-Poincare method is a technique of uniformly approximating 
periodic solutions to ordinary differential equations when regular perturbation method fails. 
This method removes secular terms without bound- arising straightforward application 
of perturbation theory to weakly non linear problems with finite oscillatory solutions.  
Lindstedt-Poincare method \cite{nayfeh1, nayfeh2} gives uniformly valid asymptotic expansion for the
 periodic solutions of weakly nonlinear 
oscillations. This method does not work for strongly non linear term. The basic idea in this method is to
 introduce a new variable. 
 \begin{equation}
 \tau = {\omega(\epsilon)} t 
 \end{equation}
 where $\omega$ is the frequency of the system. Then, any general nonlinear  oscillator equation can be written as
 \begin{equation}
 {\omega}^2 m \ddot{u} + {\omega_0}^2 u + {\epsilon} f(u) = 0
 \end{equation}
We can expand frequency in power series in $\epsilon$. Cheung et al \cite{cheung} introduced a new  parameter $\tau$. 
All these modifications are valid for conservative system. if we write the transformation as 
\[  \tau = t + f(\epsilon,t) \] 
with $ f(0,t) = 0 $ , the differential equation can be written as 
\begin{equation}
\ddot{u}  = {\left( 1 + \cfrac{\partial f}{\partial t} \right) }^2 \cfrac{\partial^2 u}{\partial \tau^2} 
= G(\epsilon, \tau) \cfrac{\partial^2 u}{\partial \tau^2} 
\end{equation}
 
The best example is unforced, undamped Duffing equation. 
A number of variants of the classical perturbation methods have been 
proposed to analyze conservative oscillators.  Burton \cite{burton} defined a parameter $ \alpha= \alpha(\epsilon) $ 
in such a way that asymptotic solutions in power series in $ \alpha$  
converges more quickly than asymptotic solution. 
Lindstedt - Poincare method gives uniformly valid asymptotic 
expansion for periodic solutions of weakly non linear oscillations. However,  this technique does not 
work in case of strongly non linear terms.  \\

The basic idea of Lindstedt - Poincare method is to introduce a new variable  in terms of oscillation parameter 
\begin{equation} 
{\tau} = {\omega}(\epsilon) t 
\end{equation} 
where $\omega$ is the  frequency of the system. Equation () then becomes 
\begin{equation} 
{\omega}^2 m u^{\prime} + {\omega_0}^2 u + {\epsilon} f(u) = 0 
\end{equation} 
Here prime denotes derivative with respect to $\tau$. 
The frequency is also expanded in powers of $\epsilon$. i.e 
\begin{equation}
\omega = \omega_0 + {\epsilon}{\omega_1} + {\epsilon}^2 {\omega_2} + \cdots 
\end{equation}

It is of interest to note that a much more accurate relation for frequency $\omega$ can be found 
by expanding $\omega^2$ in power series of $\epsilon$. 
All these modifications are limited to conservation system, instead of linear transformation. \\
We can rewrite the transformation of standard Lindstedt Poincare method in the form 
\begin{equation} 
\begin{split}
{\tau} = t + f(\epsilon, t) \\
f(0,t) = 0 
\end{split}
\end{equation} 
where $ f(\epsilon,t)$ is an unknown function(not functional). We can also see from the following
 examples that the identification of the unknown function f is much easier than that of 
 functional in Dai method \cite{dai}.  From the relation, we obtain 
 \begin{equation}
 \cfrac{\partial^2 u}{\partial t^2} = {( 1 + \cfrac{\partial f}{\partial t} )}^2 \cfrac{\partial^2 u}{\partial {\tau^2} }  
 =G \cfrac{\partial^2 u}{\partial {\tau^2} } 
 \end{equation}
 where 
\begin{equation} 
G(\epsilon,t) = {\left ( 1 + \cfrac{\partial f}{\partial t} \right) }^2 
\end{equation} 
Applying taylor series, we have, 
\begin{equation} 
G(\epsilon,t) = G_0 + {\epsilon}G_1 +  {\epsilon}^2 G_2 + \cdots
\end{equation} 
where $G_i$ i=1,2,3  $\cdots$ can be solved from equation () . 
If we simplify new parameter as 
\begin{equation}
{\tau} = f(t, \epsilon) 
\end{equation} 
with  f(0,t) as t, then 
we have 
\begin{equation}
\cfrac{\partial^2 u}{\partial t^2} = {( \cfrac{\partial f}{\partial t} )}^2 \cfrac{\partial^2 u}{\partial \tau^2} 
\end{equation}
And expanding  ${\cfrac{\partial f}{\partial t} }^2 $ in power series 
\begin{equation}
\left( \cfrac{\partial f} {\partial t} \right)^2 = 1 + {\epsilon} f_1 + {\epsilon}^2 f_2 + \cdots 
\end{equation} 
If we consider Van der Pol equation, i.e 
\begin{equation}
\ddot{u} + u - {\epsilon} (1-u^2) \dot{u} = 0
\end{equation}
With above transformation and introducing transformation variable, we can calculate period of oscillation  for $ \epsilon \ll1 $ 
very accurately as that obtained by regular perturbation method. For large $\epsilon$ , approximate 
period has same feature as exact one 
with relative error $ <  37.5 \% $. For even larger value of $\epsilon $ , modified Lindstedt-Poincare method 
gives very good result when classical method fails to work.   Parameter expansion  method \cite{ he12} developed by
 He alternatively can
 deal this type of situation very successfully.  This method does not require time transformation like 
 Lindstedt-Poincare method . 
 Main idea is to expand solution in terms of perturbation parameter.  To search for periodic solution, , we can assume 
 \begin{equation}
 u = u_0 + {\epsilon} u_1 + {\epsilon}^2 u_2 + \cdots 
\end{equation}
Implementing  this method in pendulum equation  i.e 
\begin{equation}
\ddot{u} + sin u = 0 
\end{equation}
One solves for  time period using Bessel function. We used parameter expansion method in Classical Van 
der Pol - Duffing equation 
which arises in electrical damped oscillator circuit  given by
\begin{equation}
\ddot{x} - {\mu} (1-x^2) \dot{x} + x + {\alpha} x^3 = 0 
\end{equation}
with initial boundary condition 
$ x(0) = x_0 $ and $  \dot{x(0)} = \dot{x(0) } $ Here dot denotes time derivative with $\mu$ and $\alpha$ as positive unknown 
coefficient.  Applying parameter expansion method to solve this equation 
\begin{equation} 
x = x_0 + {\epsilon} x_1 +  {\epsilon}^2 x_2 + \cdots 
\end{equation}
${\epsilon}$ is the expansion parameter . We expand the coefficients of the system in terms of $\epsilon$ 
we obtain 
\begin{equation} 
\begin{split}
{\mu} = {\epsilon} {\mu_1} + {\epsilon}^2 {\mu_2} + \cdots \\
{\alpha} = {\epsilon} {\alpha_1} + {\epsilon}^2 {\alpha}_2 + \cdots \\
1 = {\omega}^2 + {\epsilon} {\omega_1} + {\epsilon}^2 {\omega_2} + \cdots 
\end{split}
\end{equation}
$\omega$ is the frequency of non linear oscillation. Substituting these into differential equation, 
we obtain
\begin{equation}
\begin{split}
\ddot{x_0} + {\omega}^2 {x_0} = 0 \\
x = A sin \sqrt{1 + \cfrac{3}{4} {\alpha} (A^2 + B^2)} 
+B cos  \sqrt{ 1 + \cfrac{3}{4} {\alpha} (A^2 + B^2) t } 
\end{split}
\end{equation}
with frequency value obtained same as perturbation method.   Nayfeh \cite{nayfeh1} studied 
free oscillation  with non linearity of quadratic and cubic order.

\section{Perturbation Method}
Perturbation method has proved to be powerful tool in solving various boundary layer problems , interface 
flow or nano fluid mixture \cite{dutta}.
In regular perturbation problem, as $ \epsilon \rightarrow 0 $ (perturbation parameter) , the solution tends to the solution 
for $\epsilon =0 $ . Consider the equation 
\begin{equation}
\ddot{y} + 2 {\epsilon} \dot{y} - y = 0 
\end{equation}
with boundary condition $ y(0)=0$, $y(1)=1$ and $ 0 \ll \epsilon \ll 1 $. 
The general solution is 
\begin{equation} 
y(x, \epsilon) = \cfrac{ e^{m_1 x} - e^{m_2 x} } {e^{m_1} - e^{m_2} } 
\end{equation}
where $ m_1 = -  {\epsilon} + \sqrt{1 + \epsilon^2 } $  and $ m_2 = - \epsilon + \sqrt{ 1 + \epsilon^2} $ 
As $ epsilon \rightarrow 0 $, the solution becomes 
\begin{equation}
y \rightarrow \cfrac{ sinh (x) } { sinh (1) } 
\end{equation}
We can expand solution as perturbative solution. This is regular perturbation problem. 
In boundary layer theory , perturbative methods prove to be very useful tool in fluid dynamics.  
Part of the problem can be linearized as part of perturbation procedure.  The ideas stem 
from rigorous background of Liapounov - Schmidt theory of bifurcation. Falkner - Skan equation is the generalized 
equation for Blasius solution for wedge flow in fluid mechanics i.e flows are not parallel to the plate. 
In physics and fluid mechanics, Blasius boundary layer describes steady two dimensional laminar boundary layer. 
Considering Falkner - Skan equation, 
\begin{equation}
\dddot{f} + f\ddot{f} + {\beta} (1-{\dot{f}}^2) =0 
\end{equation}
with Mixed boundary condition $ f(0) = \dot{f}(0) = 0 $ and $ \dot{f}(\infty) = 1 $ to have solution 
$ f(\eta )$ for  $ 0 <   \eta   <  1 $. 
 If we expand solution in terms of perturbation parameter  as
 \begin{equation}
 f(\eta) = f_0(\eta) + ( \beta - \beta_0) f_1(\eta) + ( \beta - \beta_0)^2 f_2 (\eta) + \cdots 
 \end{equation}
 On equating coefficients of powers of small parameter $ \beta - \beta_0$ in
  Falkner Skan equation, this gives us a succession of linear inhomogeneous problems to solve 
  \begin{equation} 
  \begin{split}
  Lf_1 = \dddot{f}_1 + (\beta-\beta_0) f_1(\eta) + (\beta-\beta_0)^2 f_2(\eta) + \cdots \\
  Lf_2 = {\beta_0} {\dot{f}_1}^2 - {f_1} \ddot{f}_1 + 2 \dot{f}_0 \dot{f}_1 
  \end{split}
  \end{equation} 
 We can implement perturbation method to find solution.  And  we obtain solution of following form 
 \begin{equation}
 f(\eta) = f_0(\eta) \pm { (\beta - \beta_0) ( I_1/I_2 ) }^{1/2} {f_0(\eta)}^{\prime} + O(\beta - \beta_{\star} ) 
 \end{equation}
 with existence of bifurcation near $ \beta = \beta_0$. 
The Blasius boundary layer solution is only one member of a larger family of boundary
 layer solution known as Falkner  Skan flow.  Considering Blasius boundary layer equation applied to two 
 dimensional flow over a wedge with slip flow boundary, one can implement perturbative method very 
 successfully in numerical analysis \cite{martin}. 

\section{Iterative Perturbation Method} 
In this section, we will illustrate a new perturbation technique 
coupling with the iterative method. Consider the following nonlinear 
oscillation: 
\begin{equation}
 u^{\prime} + u + {\epsilon} f(u,u^{\prime}) =0 
\end{equation}
We write the equation in the following form 
\begin{equation} 
 u^{\prime} + u + {\epsilon}u g(u,u^{\prime}) = 0 
\end{equation}
We construct an iteration formula for the above equation 
\begin{equation}
 {u_n+1}^{\prime} + u_{n+1} + {\epsilon}u_{n+1} g(u_n, {u_n}^{\prime}) =0 
\end{equation}
For nonlinear oscillation, Eq () is of Mathieu type. We will use 
the perturbation method to find approximately $u_{n+1}$. 
Consider Eq () and assume that initial approximate solution can be
 expressed as $ u_0 = A cos {\omega}t $ where $\omega$ is the angular
frequency of oscillation. We rewrite Eq () approximately 
as follows 
\begin{equation}
\cfrac{d^2 u}{d t^2} + {\epsilon} A^2 u  {cos \omega t}^2 = 0
\end{equation}
or 
\begin{equation}
  \cfrac{d^2 u}{dt^2} + \cfrac{1}{2}{A^2 u }  {\epsilon} {A^2 u }  {cos 2 {\omega}  t} = 0
\end{equation}
which is Mathieu type. Suppose that 
\begin{equation}
 u=u_0 + {\epsilon} u_1 + {\epsilon}^2 u_2 
\end{equation}
To obtain a second order approximate solution, we substitute $u_0$ and 
$u_1$ we find
\begin{equation}
 c_2 = \cfrac{c_1 A^2}{32 \omega^2} 
\end{equation}
To obtain approximate solution with higher accuracy, we

 \section{Parametrized Perturbation Method} 
 An expanding parameter is introduced by a linear transformation 
 \begin{equation} 
 u = {\epsilon}v + b 
 \end{equation} 
 where $\epsilon$ is the introduced perturbation parameter, b is a constant. 
 
 \section{Multiple Scale  Lindstedt Poincare Method(MMS)}
 
The method of Multiple Scale (MMS) is routinely applied to many weakly nonlinear systems
of both ordinary and partial differential equations.  For delay differential equations or retarded functional 
differential equations in the neighborhood of Hopf bifurcation , this method had been applied very 
successfully without applying manifold reduction scheme. 
Manifold reduction as suggested by many authors \cite{cambell} eliminates a priori many exponentially 
decaying part of the solution leaving behind slowly evolving dynamics on the center manifold.  
MMS method can efficiently 
eliminate rapidly decaying part in the center manifold.  This method is very efficient in system with  time delay 
\cite{plaut}.  If we consider autonomous equation with limit cycle 
\begin{equation} 
\dot{y}(t) = - {\alpha} y (t-\cfrac{\pi}{2} ) - y^3 (t) 
\end{equation}
with $\alpha >0$ . The characteristic equation for the linearized form is 
\begin{equation}
\tilde{\lambda} e^{\tilde{\lambda}} + \tilde{\alpha} = 0 
\end{equation}
For $ 0 < \tilde{\alpha} < \cfrac{\pi}{2} $ , this equation undergoes Hopf bifurcation 
at $  \tilde{\alpha} = \cfrac{\pi}{2} $ or $\alpha =1 $. To study equation near bifurcation, 
we take $ \alpha = 1 + \epsilon$ and equation () becomes 
\begin{equation}
\dot{y}(t) = - (1 + \epsilon) y (t - \cfrac{\pi}{2} ) - y^3 (t) 
\end{equation}
with $ 0 < \epsilon \ll 1 $.  If we allow transformation  solution $ y = {\epsilon} x $ , equation () becomes 
\begin{equation}
\dot{x}(t) = -x (t-\cfrac{\pi}{2}) - {\epsilon} [ x(t-\cfrac{\pi}{2}) + x^3 (t) ] 
\end{equation} 
Hopf bifurcation occurs at $ \epsilon=0$ and a stable periodic solution exists  for small $ \epsilon > 0$.  
To find solution near bifurcation point, we can define multiple time scale to implement MMS of the form
\begin{equation}
x(t) = X_0(t,T_0,T_1,T_2)  + {\epsilon} X_1(t,T_0,T_1,T_2)  + {\epsilon}^2 X_2(t,T_0,T_1,T_2) + \cdots 
\end{equation}
 Substituting the above solution in the differential equation, we obtain
 \begin{equation}
 \cfrac{\partial X_0} {\partial t} + {\epsilon} \left(  \cfrac{\partial X_1} {\partial t} +  \cfrac{\partial X_0} {\partial t} 
 - \cfrac{\pi}{2}  \cfrac{\partial X_0} {\partial t} (t- \cfrac{\pi}{2}) \right) = 
 - X_0 ( t- \cfrac{\pi}{2}) - {\epsilon} (t- \cfrac{\pi}{2}) + X_0 (t-\cfrac{\pi}{2}) + {X_0}^3(t)] + O({\epsilon}^2) 
 \end{equation}
 This equation has of the form 
 \begin{equation}
 X_0 = A sin t + B cos t 
 \end{equation}
And if one solves for  solution , one observes MMS result is very close to Direct Numerical Simulation result for oscillation. 
MMS method can be implemented  to strongly nonlinear  case where the method needs to be carefully implemented. 
 \section{Homotopy Perturbation Method} 
 Homotopy Analysis method (HAM) is a semi analytical technique to solve 
 non linear/partial differential equations. The homotopy analysis method employs 
 the concept of homotopy from topology  to generate a convergent series solution 
 for non linear systems.  This is enabled by utilizing  a homotopy - Maclaurin series 
 to deal with the non linearity of the system. 
 Homotopy perturbation method have been applied to various kinds of non linear problems. 
 The main property of the method is its flexibility and ability to solve non linear equations 
 accurately 
 and conveniently used.  Since perturbation method depends on small parameter which is very difficult 
 to be found in non linear problems, Variational Perturbation Method(VPM)  and Homotopy Perturbation Method(HPM)
  have been introduced 
 to solve non linear problems.   HPM method  is being as most effective method in non linear 
 analysis of engineering problems.  
 In this method, according to homotopy technique, a homotopy with embedded parameter 
 $ p \in [0,1] $ is constructed , so the parameter embedded is small , so the method is called homotopy 
 perturbation method.  This method can take the full advantage of homotopy techniques and perturbation method. 
 This method is widely used in Duffing equation with high order of non linearity. 
 Consider a non linear equation  in the form 
 \begin{equation} 
 A(u) + f(r)  = 0 
 \end{equation}
 where $ r \in \Omega $ with boundary conditions  
 \begin{equation} 
 B(u, \cfrac{\partial u}{\partial n} ) = 0 
 \end{equation} 
 with $ r \in \Gamma $. 
 where A is general operator and B is boundary operator. f(r) is a known analytic function and $\Gamma $ is t
 he boundary of the domain $
 \Omega$. The operator , in general be divided  into two parts L and N, where L is the linear and M is the non linear part. Equation() 
 can be written as 
 \begin{equation}
 L(u) + N(u) - f(r) = 0
 \end{equation} 
 In case of non linear equation, it has no small parameter. We can embed an artificial parameter
  in the above equation as 
 \begin{equation}
 L(u) + pN(u) -f(r) = 0 
\end{equation} 
 In this method, according to homotopy technique, a homotopy with embedded parameter 
 $ p \in [0,1] $ is constructed , so the parameter embedded is small . The above equation is called 
 perturbation equation 
 with artificial small parameter 
 p and it assumes that u can be expressed as a series of power of p. 
 \begin{equation}
 u = u_0 + pu_1 + p^2 u-2 + p^3 u_3 + \cdots 
 \end{equation} 
  In order to use 
 homotopy perturbation,  a suitable construction of homotopy equation is of vital importance. 
 For example , \\
 \begin{equation} 
 y^{\prime} + y^2 = 1 
 \end{equation} 
 with $ t > 0$ with initial condition with initial condition $ y(0)=0$. We embed an artificial parameter p such that 
 \begin{equation} 
 y^{\prime} + p y^2 = 1
 \end{equation}
 This equation does not have approximate solution in power of p. So,  Homotopy technique 
 proposed by Liu et al \cite{liu} 
 is being applied here which satisfies 
 \begin{equation} 
 (1-p)[ L(v) - L(u_0) ] + p[A(v) -f(r) ] = 0 
 \end{equation} 
 which satisfies the boundary condition.  
 Transformation of p from zero to unity is equivalent to deformation in topology and 
 $ L(v) - L(u_0)  and A(v) -f(r)  $ are homotopic. 
 v(r,p) changes from $  u_0 to u(r) $  homotopically. 
 It is very natural to express v as 
 \begin{equation}
 v = v_0 + pv_1 +{ p^2 } { v_2 } + c\dots 
 \end{equation}
 In the limit $ p \rightarrow 1 $, 
 \begin{equation}
 u = \lim p \rightarrow 1  v = v-0 + v_1 + v_2 + \cdots 
 \end{equation} 
 Convergence of this series has been proved in \cite{he3}, \cite{he4}. 
  Using method of weighted residual methods, one can obtain  p value from the above equations. 
  We cite some examples. Consider very simple system 
  \begin{equation} 
  y^{\prime} + y^2 = 0 
  \end{equation} 
  $ x \ge 0$ , $ x\in \Omega $ with boundary condition y(0)=1. 
  with exact solution . \\
  We construct solution following homotopy $ \Omega \times [0,1] $ which satisfies 
  \begin{equation} 
  (1-p)(y^{\prime} - y_0^{\prime}) + p( y^{\prime} + y^2 ) = 0 
  \end{equation} 
  with initial approximation $ y_0 =1 $. Suppose the solution has the form 
  \begin{equation} 
  y = y_0 + py_1 + p^2 y_2 + \cdots 
  \end{equation} 
  Now we consider a typical non linear equation, i.e Duffing equation with non linearity 
  with higher order non linearity . Duffing equation represents simple harmonic motion of 
  damped pendulum in may cases. 
  \begin{equation}
  \cfrac{d^2 u}{dt^2} + u + {\epsilon} u^5 = 0 
  \end{equation} 
  $ t \in \Omega $ with initial condition  $u(0)= u^{\prime} (0) = 0 $. 
  we construct a homotopy  $ \Omega \times [0,1] $ which satisfies 
  \begin{equation} 
  L(v) - L(u_0) + p L(u_0) + p {\epsilon} v^5 = 0 
  \end{equation} 
  where $ Lu =   \cfrac{d^2 u}{dt^2} + u $. We assume initial guess of equation () as 
  \begin{equation}
  u_0(t) = A \ cos {\alpha} t 
  \end{equation} 
  where $ \alpha(\epsilon) $ is a unknown constant with $ {\alpha} (0) = 1 $. \\ 
  with similar approximate solution for v we have, 
  \begin{equation} 
  \begin{split}
  L(v_0) - L(u_0) = 0 \\ 
  L(v_1) + L(u_0) + {\epsilon}{v_0}^5 = 0 
  \end{split}
  \end{equation}
  with $ {v_1}^{\prime} (0) = v_1(0) = 0 $. 
  From equation () and () we obtain, 
  \begin{equation}
  \begin{split} 
v_0 = A \ cos {\alpha} t \\
\cfrac{d^2 v_1}{d t^2} + v_1 + A \left( - {\alpha}^2 + 1 + \cfrac{5}{8} {\epsilon} A^4 \right) \cos {\alpha} t 
+ \cfrac{\epsilon A^5}{16} ( cos 5 {\alpha}t + 5 \ cos 3 {\alpha}t ) 
\end{split}
\end{equation} 
The solution of the above equation can be obtained using variational Iteration method. i.e 
\begin{equation}
 v_(t) = {\int_0}^t \ sin(t-\tau) [ A ( -{\alpha}62 =1 + \cfrac{5}{8} {\epsilon} A^4 ) \ cos {\alpha} {\tau}  
\end{equation}
 The constant $\alpha$ can be identified as weighted residual (least square method, collocation method, 
 Galerkin method). 
 So, we found $v_1(t) $ as 
 \begin{equation}
v _1(t) = \left(  - {\alpha}^2 + 1 + \cfrac{5}{8} {\epsilon} A^4 \right) \cfrac{A}{\alpha^2 -1} \ cos {\alpha} t + 
 \cfrac{\epsilon A^5} {16(25 \alpha^2 -1) } \ cos 5 {\alpha}  t + \cfrac{5 \epsilon A^5}{16(9 \alpha^2 -1)} 
 \end{equation}
 Finally we calculate $u_1(t)$ 
 \begin{equation} 
 \begin{split}
 u_1(t) = v_0(t) + v_1(t) \\
 u_1(t) = \cfrac{5 \epsilon A^5 }{8(\alpha^2-1)} \ cos \alpha t + \cfrac{\epsilon A^5}{16( 25 \alpha^2 -1)}
 \cos 5 \alpha t + c\frac{ 5 \epsilon A^5}{16(9 \alpha^2 -1) } cos 3 {\alpha}  t 
 \end{split}
 \end{equation} 
 Nofel et al \cite{nofel} implemented homotopy perturbation method in obtaining approximate periodic solutions for 
 non linear heat conduction equation and Van der Pol equation. 
 Biazar et al \cite{biazar} implemented homotopy perturbation method in detail  for solving
  hyperbolic equations. Hyperbolic partial differential equations play a dominant role 
  in many branches of science in engineering. Wave equation is typical example , which models problems 
  such as vibrating string fixed at both ends , longitudinal vibration in bar, propagation of sound waves, 
  electromagnetic waves. They also occur in fluid mechanics.  Euler equations in gas dynamics are 
  typical examples of hyperbolic equations.  They considered following homotopy equations 
  for the quasi linear hyperbolic partial differential equation
  \begin{equation} 
 L(u) + N(u) - f(r) = 0 
  \end{equation} 
 Using homotopy technique, one can construct homotopy equation as
 \begin{equation} 
 H(u,p) = (1-p) 
 \end{equation} 
 $p \in [0,1] $ is the homotopy parameter with $u_0$ as initial guess satisfying boundary conditions. 
 Considering the problem of combined convective -radiative cooling of 
 lumped system represented by equation.
 
 \section{Singular Perturbation Method}
Singular perturbation problems often arise in fluid mechanics with thin boundary layer , the properties of slightly viscous fluid
 dramatically changes on both sides of boundary layer so that fluid exhibits multiple spatial scales. Reaction diffusion systems
 are classic example of this system. 
 A wide variety of techniques have been applied to solve singular perturbation problem. In one form or other, 
 these techniques consist of dividing the problem into an inner solution(or boundary layer) and an outer layer problem. 
 Express the inner and outer as asymptotic series , equating various terms on inner and outer solutions in the series. 
 Finally, one obtains solution combining the inner and outer solutions in some fashion to obtain uniformly valid solution. 
 Commonly the outer 
 solution is obtained as a function of small(perturbation) parameter with $\epsilon \rightarrow 0$.  Kaplun \cite{kaplun} first
  developed ideas about singular 
 perturbation in studying asymptotic solution of Navier Stokes equation for low Reynold number based on limit expansion idea. 
 Under outer limit, finite body shrinks to a point which can not disturb the flow or presence  of small body causes perturbation 
 which is small everywhere except surface. 
 The inner limit is the one where viscosity tends to infinity.  Mathematical technique for constructing asymptotic 
 expansion consists in first
  finding the leading term of outer solution and matching inner solution to outer approximation.  Kaplun considered 
  partially ordered set of equivalence classes of functions f(Re). To each such f(Re), corresponding limit process  is 
  defined and implemented in N-S equations. 
 Boundary value technique is being introduced by Roberts \cite{roberts} to obtain asymptotic 
 matched coefficients in various situations.  Boundary technique partitions original singular perturbation equation into
  two problems , 
 an inner solution differential equation and an outer solution differential equation.  The solution of the outer equation 
 provides  
 provides terminal conditions for inner equation. And in turn, solution for inner equation provides initial condition for
  outer equation. 
 The problem is solved iteratively for various values of terminal point and terminal boundary conditions of inner
  solution until profile 
 stabilizes and original boundary condition is satisfied. 
  Consider following singular perturbation equation 
 \begin{equation}
 {\epsilon} y^{"} (x) + y^{\prime}(x) + y(x) = 0 
 \end{equation}
 with boundary condition $ y(0)= \alpha$ , $ y(1) = \beta$ and $ 0 \le x \le 1 $. 
  Converting  the original singular perturbation 
 equation to the outer solution with right boundary condition, outer equation reduces to 
 \begin{equation} 
 y^{\prime}(x) + y(x) = 0 
 \end{equation}
 with  $ y(1) = \beta$. Parameter scaling is very important to develop inner differential equation.
  Here we set $ t= \cfrac{x}{\epsilon} $ and equation becomes 
  \begin{equation}
  \begin{split}
  y(x) = Y(t) \\
  y^"(x) = \cfrac{Y^{\prime} (t)}{\epsilon^2} 
 \end{split}
 \end{equation}
   to obtain differential equation of inner solution 
  \begin{equation}
  Y^"(t) + y^{\prime} (t) + {\epsilon} Y(t) = 0 
  \end{equation} 
  with boundary condition .  choosing $t_f$ as terminal point for inner solution, one obtains inner equation as
   initial value problem 
  with 
  \begin{equation}
  y(t_f \epsilon) = Y(t_f) = \tilde{\beta} 
  \end{equation}
  The original boundary value perturbation equation becomes initial value equation and can be solved numerically iteratively. 
Singular perturbation problem is one where perturbed problem is qualitatively different from unperturbed one. 
 Solution of singular perturbation problems involving differential
 equations often depend on several widely different length or time scale. Such problems can be divided
  into two 
 broad classes: layer problem being treated using method of matched asymptotic expansions (MMAE) and  multiple scale
  problem treated
  by method of  multiple scale  (MMS). Prandtl boundary layer theory for high Reynold number flow of viscous fluid 
  over a solid body  is classic example of MME method and semi classical limit of quantum mechanics is an example of 
   MMS problem.  An asymptotic expansion as $ z \rightarrow \infty$ of a complex function with essential singularity
 is valid only in wedge shaped region.  MMS method is needed for problems in which the solutions depend 
 simultaneously on widely different scales. A typical example is the modulation of oscillatory solution over
  time scale that is much greater than period of oscillation.  Considering boundary layer problem in fluid mechanics , 
   Boundary layer theory and asymptotic matching are collection 
   of singular perturbation methods for constructing a uniformly and globally valid solution by calculating 
   the separated outer and inner solutions and matching them at boundary.     Considering non linear boundary layer problem . 
   
 For example , Navier Stokes equation can be expanded  ( a highly non linear equation) 
 in asymptotic power series 
 in small parameter $\epsilon$ which is determined as a function of Reynold number by asymptotic technique. 
 Flow over constant pressure flat plate turbulent boundary layer had been studied by asymptotic technique \cite{yajnik}, 
 \cite{ meller}.  Starting with two dimensional Navier Stokes equation in asymptotic power series of $\epsilon$ in 
 both region
 i.e inner and outer region and then matching both equations in the overlap region in asymptotic manner. 
 We assume, asymptotic parameter  $\epsilon$ as function of Reynold number and goes to zero as 
 Reynold number $\rightarrow  \infty$. And we can write velocity field in Navier Stokes equation
 as 
 \begin{equation} 
 U = \tilde{i} + {\epsilon} U_1(X-T,Y,Z) + \cdots 
 \end{equation}
 Matching outer and inner equations at boundary , Meller et al \cite{meller}  developed 
 asymptotic expansion identity for velocity field 
 \begin{equation}
 U = {\epsilon} \left( \kappa^{-1} ln ({\epsilon} R_e) + B_i  \right)
 \end{equation}
 Meller showed in their asymptotic analysis that overlapping zone exists in solution and it indicates coupling
  between inner and outer 
 layer at the turbulent boundary layer.  Method of matched asymptotic expansion is particularly
  used for singular perturbed differential equation. 
   
  Fundamental idea of asymptotic ( or perturbative) analysis is to relate unknown solution of original complex 
  problem to the known simple solution using simple transition. 
  Consider a general equation, 
  \begin{equation}
  L(u) + {\epsilon} N(u) = 0 
  \end{equation} 
  where $ L(u)=0$ is an equation that can be readily solved with the solution $u_0$. 
  We assume N is a bounded operator defined on a space in real valued function. 
  We need to find $ u({\epsilon}) $ to above equation with condition 
  $ u(\epsilon) \rightarrow 0 $ as $ \epsilon \rightarrow 0$. And since L(u) is a linear operator 
  \begin{equation}
  L(c-1u + c_2v) = c_1 L(u) + c_2 L(v) 
  \end{equation}
  holds and the solution can be found as 
  \begin{equation}
  u(\epsilon) = u-0 + {\epsilon} u_1 + {\epsilon}^2 u_2 + \cdots 
  \end{equation}
  This is regular perturbation method.  For singular perturbed boundary problem, 
  For boundary problem, we employ asymptotic expansion(outer and inner ). 
    We will use MMS method to construct leading asymptotic terms of the outer and inner expansion
  of the solution to the singularly perturbed solution.  Passing to the limit, as $ \epsilon \rightarrow 0$, 
  we observe that Dirichlet boundary condition disappear and we arrive to following limit. 
  All non trivial solutions have singularities and behavior of functions can be analyzed through 
  matching of the outer with inner at the overlap region. To construct inner asymptotic 
  expansion of the solution, generally we pass through the stretched coordinates $ \xi^j $
   and construct as
   \begin{equation}
   u^{\epsilon}(x) = {\mu_0}(\epsilon) {w_0}^j{(\xi^j} + {\mu_1}(\xi) {w_1}^j ( \xi^j) + \cdots 
   \end{equation}
   To calculate asymptotic series outer expansion , we construct leading asymptotic term first 
\begin{equation}
{w_0}^j(\xi) = {\xi} 
\end{equation}

\subsection{WKB Method}
WKB theory  is powerful tool to solve 
   linear system to find global uniform solution and can be applied to boundary layer problem. This  method is  
   generally applied  in solving linear differential equation and can be implemented 
effectively in case of non linear equations also. WKB method is a method of 
approximating the solution of a differential equation whose highest derivative is multiplied 
 by a small parameter $\epsilon$.
Miura  \& Kruskal  \cite{miura} first implemented WKB method 
effectively to solve kdV equation in presence of small dispersion parameter.    
Let us consider general differential equation
\begin{equation}
{\epsilon}^2 \cfrac{d^2 y}{d x^2} + F(x) y =0 
\end{equation} using WKB method. 
We can assume the solution in the form
\begin{equation}
y(x, \epsilon) = A(x,\epsilon) exp\left( \cfrac{iu(x)}{\epsilon} \right) 
\end{equation}
  so that differential equation becomes 
  \begin{equation}
  (-({u^\prime}^2 ) A + {\epsilon} i( A u^{"} + 2 A^{\prime} u^{\prime} ) + {\epsilon}^2 A^{"} ) 
 \ exp(\cfrac{i u}{\epsilon} + F(x) exp(\cfrac{iu}{\epsilon} = 0 
  \end{equation}
  Now let 
  \begin{equation}
 A(x,epsilon) \approx A_0(x) + {\epsilon}A_1(x) + \cdots 
  \end{equation} 
  And the solution just comes as 
  \begin{equation}
  u(x) = \pm  {\int_{x_0}}^x \sqrt{F(s) } ds 
  \end{equation} 
  WKB method can be implemented in nth order differential equation \cite{bender, takei}. One of the most common 
  application of WKB
   method occurs in  solving Schroedinger equation in quantum mechanics. WKB method is also  used in geometrical optics,
   Ray theory, various wave patterns specially with Kelvin ship wave.  WKB method can be used in solving 
   Schroedinger- Newton equation which form a class of non linear differential equations. 
   These equations are obtained by coupling together the ordinary Schroedinger equation and Newton field of equations 
   for the gravitational potential. The equations describe a particle moving in its own gravitational field , where the field is 
   generated via classical Newtons field equation. This equation is originally obtained by Penrose \cite{pen} .
   Considering time independent equations of the form 
   \begin{equation}
   \begin{split} 
   \Delta \psi = - \cfrac{2m}{\hbar^2} (e-\phi) {\psi} \\
   \Delta \phi = - 4 {\pi} G m^2 \mid \psi\mid^2 
   \end{split}
   \end{equation} 
   where $\psi$ is the wave function and $\phi$ denotes grav. potential. E is the energy eigen value.  If we define variables as
   $ U= \cfrac{2m}{\hbar} (E-\phi) $ and $ S= \cfrac{\hbar \psi}{\sqrt{  8 pi G m^3}} $, one may evolve
    equations 
    \begin{equation}
    \begin{split} 
    \Delta S = -SU \\
    \Delta U = - S^2 
    \end{split}
    \end{equation}
    These equations are known as Schroedinger Newton equation. Implementing  WKB method to these equations, one obtains 
    \begin{equation} 
    \begin{split}
    S(r) = S_0 - {\int_0}^r x(1-\cfrac{x}{r} ) S(x) U(x) dx \\
    U(r) = U_0 - {\int_0}^x  x(1-\cfrac{x}{r})  S(x)^2 dx 
    \end{split} 
    \end{equation} 
  
  In nanoscale semi conductor structure studies,  quantum effects occur and have to be taken into account 
  by means of Schroedinger equation. Computational difficulties develop in numerical calculation of Schroedinger equation. 
  On the other hand 
  behaviour of the solution requires a refined spatial grid. Wang et al \cite{wang2} 
   developed WKB-LDG method  in simulating resonant tunneling diode 
  which provides computational and memory cost significantly.

 \end{document}